\documentclass[twocolumn,secnumarabic,amssymb, nobibnotes, aps, prd]{revtex4-2}

\usepackage{graphicx, bm}
\begin{document}

\title{A pre-screening method for variational quantum state eigensolver}

\author{Hikaru Wakaura}%
\email[Quantscape: ]{
hikaruwakaura@gmail.com}
\affiliation{QuantScape Inc., 4-11-18, Manshon-Shimizudai, Meguro, Tokyo, 153-0064, Japan}

\author{Andriyan B. Suksmono}

\affiliation{ Institut Teknologi Bandung, Jl. Ganesha No.10, Bandung, Jawa Barat, Indonesia}
\email[Bandung Institute of Technology: ]{suksmono@stei.itb.ac.id}

\date{\today}%

\begin{abstract}
The development of Fault-Tolerant Quantum Computer (FTQC) gradually raises a possibility to implement the Quantum Phase Estimation (QPE) algorithm. However, QPE works only for normalized systems. This requires the minimum and maximum of eigenvalues of a Hamiltonian. Variational Quantum Eigensolver (VQE) well developed in the Noisy Intermediate Scale Quantum (NISQ) era is necessary for preparing the initial eigenvectors that close to the exact states. In this paper, we propose a method to derive all of the states with high accuracy by using the Variational Quantum State Eigensolver (VQSE) and Subspace-Search VQE (SSVQE) methods. We show that by using the  VQSE and the SSVQE prescreening methods, we can derive all of the hydrogen molecules states correctly.
\newline

\end{abstract}

\maketitle

\section{Introduction}\label{1}

Since quantum computer was proposed by Feynman\cite{feynman_simulating_1982}\cite{2021arXiv210610522P}, various algorithms that take advantage of the quantum computers have been developed; among others are the Grover's\cite{2003quant.ph..1079L}, Shor's\cite{365700}, and Quantum Phase Estimation (QPE) algorithms\cite{1995quant.ph.11026K}. Especially, the QPE can be used to derive the eigenstates and their eigenvalues of Hamiltonian systems by using a quantum computer. Unfortunately, these algorithms cannot be performed without Quantum Error Correction (QEC)\cite{2019ConPh..60..226R}\cite{chen_exponential_2021}. At present, quantum computers are in its NISQ (Noisy Intermediate Scale Quantum) era, which means that their qubit numbers are  too few to perform the QEC.

Nevertheless, various algorithms for the NISQ devices have been developed. The Variational Quantum Eigensolver (VQE) introduced by Aspuru-Guzik et al. in 2011 \cite{doi:10.1146/annurev-physchem-032210-103512} has been investigated and improved by various groups. Today, various kinds of the VQE methods are emerges. For example, by using a Subspace-Search VQE (SSVQE) \cite{PhysRevResearch.1.033062}, we can calculate multiple energy levels at once, wheras the Multiscale-Contracted VQE (MCVQE) \cite{2019PhRvL.122w0401P} enables calculation of the ground and single electron excited states by diagonalizing the configuration interaction state of a Hamiltonian. Adaptive VQE \cite{2019NatCo..10.3007G} and Deep-VQE \cite{2020arXiv200710917F} are also proposed and even the essential procedure of the VQE has been exploited for a quantum version of machine learning \cite{PhysRevA.98.032309}.

Some research groups are close to achieve building a Fault-Tolerant Quantum Computer (FTQC)\cite{Gambetta2017}. Hence, quantum algorithms for the FTQC era will be used in not so distance future. The QPE itsels also have been developed for a while\cite{PhysRevA.76.030306}\cite{2020PhRvA.102b2422P}. However, this algorithm requires the normalization by the minimum and maximum of the Hamiltonian's eigenvalues which should be close to the exact states. Hence, some kind of VQE algorithm should be developed for the era of FTQC to assist the QPE.

Variational Quantum State Eigensolver (VQSE)\cite{2020arXiv200401372C} is kind of VQE advanced method that optimizes all of the eigenvectors and their eigenvalues using a density operator. Nevertheless, VQSE cannot be used to derive all the eigenstates when all energy levels are too distant and some of them degenerate. A pre-screening or similar method is necessary, such as the  Moller-Plesset 2 (MP2) method\cite{2017arXiv170102691R}. To address this issue, we propose a new pre-screening method for the VQSE i.e., the SSVQE-prescreening. In essence, it is a method that uses the result of the SSVQE as an initial parameter set of the VQSE. As a result, the SSVQE-prescreening is confirmed in assisting the convergence of all of the states with sufficient iteration times.

The rest of this paper are organized as follows. In Chapter \ref{2}, we describe the idea of VQSE and SSVQE-prescreening methods. In chapter \ref{3}, we compare the results of VQSE with SSVQE-prescreening and VQSE with no prescreening. In Chapter \ref{4}, we compare the convergence VQSE with SSVQE-prescreening on hydrogen and lithium hydride molecules. Chapter \ref{5} will conclude this work.

\section{Methods}\label{2}
In this section, we describe the principles of VQSE and VQSE with SSVQE-prescreening. VQSE optimizes the manifold of Hamiltonian's eigenstates using the density operators in an evaluation functions. It is expressed as follows, 

\begin{equation}
C(\bm{\theta})=Tr(HU^\dagger(\bm{\theta})\rho U(\bm{\theta})),\label{C}
\end{equation}

where $U(\bm{\theta})$ is the operator to make the given superposition state that includes the trotterized Hamiltonian and cluster terms correspond to  $\bm{\theta}$, which is a variable vector. Then, $\rho=\sum_j\lambda_j\mid \lambda_j \rangle \langle \lambda_j \mid$ is the density matrix that satisfy the condition $\lambda_{k}\leq \lambda_j, k>j$. $\lambda_j$ and $\mid \lambda_j \rangle$ are the eigenvalue and its eigenvector of given matrix. Throughout this paper, we assume that $\rho=\sum_j (N-j)/(N+1) \mid j \rangle \langle j \mid$, where $N$ is the number of the states on quantum resources and $\mid j\rangle$ is the qubit state represented by a decimal number, respectively. Assuming $\mid j \rangle$ as $\mid \Phi_{ini}^j \rangle $,  the Eq.\ref{C} can be expressed by tracing  $U(\bm{\theta})\mid \Phi_{ini}^j \rangle$ as follows,

\begin{equation}
C(\bm{\theta})=\sum_{j,k=0}^{N}\langle \Phi_{ini}^j \mid U^\dagger(\bm{\theta})HU\mid \Phi_{ini}^k \rangle \langle \Phi_{ini}^k \mid U^\dagger U\mid \Phi_{ini}^j \rangle.\label{Cp}
\end{equation}

Since the initial states are orthogonal, the off-diagonal terms of the trial energy can be omitted. Eq. \ref{Cp} can be simplified into,

\begin{equation}
  C(\bm{\theta})=\sum_{j=0}^{N}E_j p_j\label{Cm}
\end{equation}. 

The depth of Hamiltonian and cluster terms are set to $n$, which in the following discussions is equal to 2. It is expressed as $U(\bm{\theta})=\prod_l \prod_k exp(-i\theta_l c_k^lP_k^l)$ by the index of the variable $l$ and the index of terms belongs to variable $l$ transformed into Pauli words $k$ \cite{doi:10.1021/acs.jctc.8b00450} \cite{2017arXiv171007629M}. We use the Unitary Coupled Cluster(UCC) \cite{2018PhRvA..98b2322B} ansatz as a cluster. 

We propose the SSVQE-pre-screening as a new way of pre-screening steps to prepare the initial value of the parameter. This pre-screening must satisfy three conditions: (1) the basis used for the SSVQE method is the partial set of the initial basis used for VQSE, (2) the number of variables and clusters must be the same as that of VQSE, (3) the number of iteration must be sufficient to derive some states accurately. SSVQE is an advanced method of VQE that can derive multiple states at once. This algorithm minimizes the sum of evaluation functions of each state. 

\begin{equation}
F(\bm{\theta})=\sum_j\lambda_jF_j
\end{equation}

where $\lambda_j$ is the same constant as that of VQSE. The function to be minimized in actual SSVQE method is a function that contains constraint terms \cite{doi:10.1021/acs.jctc.8b00943} $E^{const}_j(\bm{\theta})$ and deflation terms of Variational Quantum Deflation(VQD) \cite{2018arXiv180508138H} $E^{def}_j(\bm{\theta})$. The method to calculate the product between two states is SWAP-test \cite{2013PhRvA..87e2330G}, which is used to calculate the excited states. The evaluation function of $j^{th}$ state is,

\begin{equation}
 (\bm{\theta})=E_j(\bm{\theta})+E^{const}_j(\bm{\theta})+E^{def}_j(\bm{\theta})\label{F}.
\end{equation}

The constraint and deflation terms are zero at a global minimum of the evaluation function. In this paper, we take the initial states of SSVQE-prescreening as $\mid 1000 \rangle, \mid 1100 \rangle, \mid 0110 \rangle$, and $\mid 0010 \rangle$ in a binary representation on both of hydrogen and lithium hydride molecules, respectively.

The Hamiltonian and the basis for the molecules are STO-3G and the depths of the cluster and the Hamiltonian are both equal to 2. All results of quantum calculations are in the form of the state vectors (with the number of shots are assumed to be infinity). The actual calculations are performed using openfermion\cite{2017arXiv171007629M} and blueqat SDK\cite{blueqat}, which are quantum computer simulators.

\section{VQSE with SSVQE-prescreening}\label{3}
\begin{figure*}[t]
\includegraphics[scale=0.60]{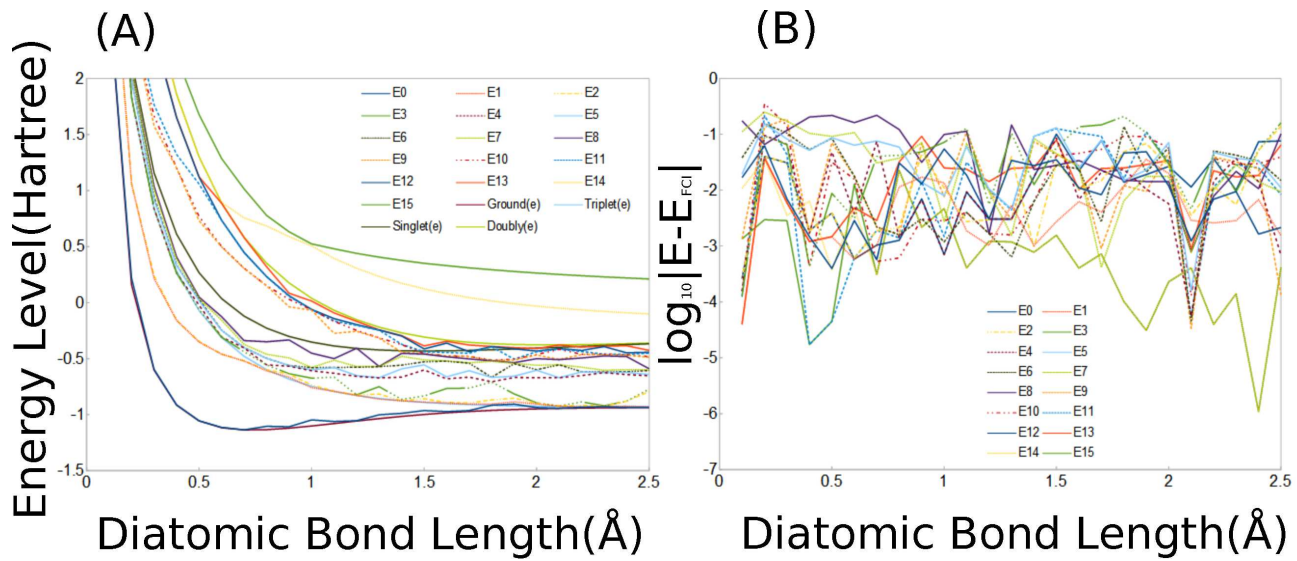}
\newline
\caption{Calculated energy levels(Hartree) of all 16 states on hydrogen molecule and their log error by VQSE with no prescreening, respectively. (A) energy levels of each state for diatomic bond length r$(\AA)$. (B) log error($log_{10}\mid E - E_{FCI} \mid$) of each states. All data are sampled from 0.1 to 2.5 in 0.1 pitch.
}\label{VQSE}
\end{figure*}

\begin{figure*}[t]
\includegraphics[scale=0.60]{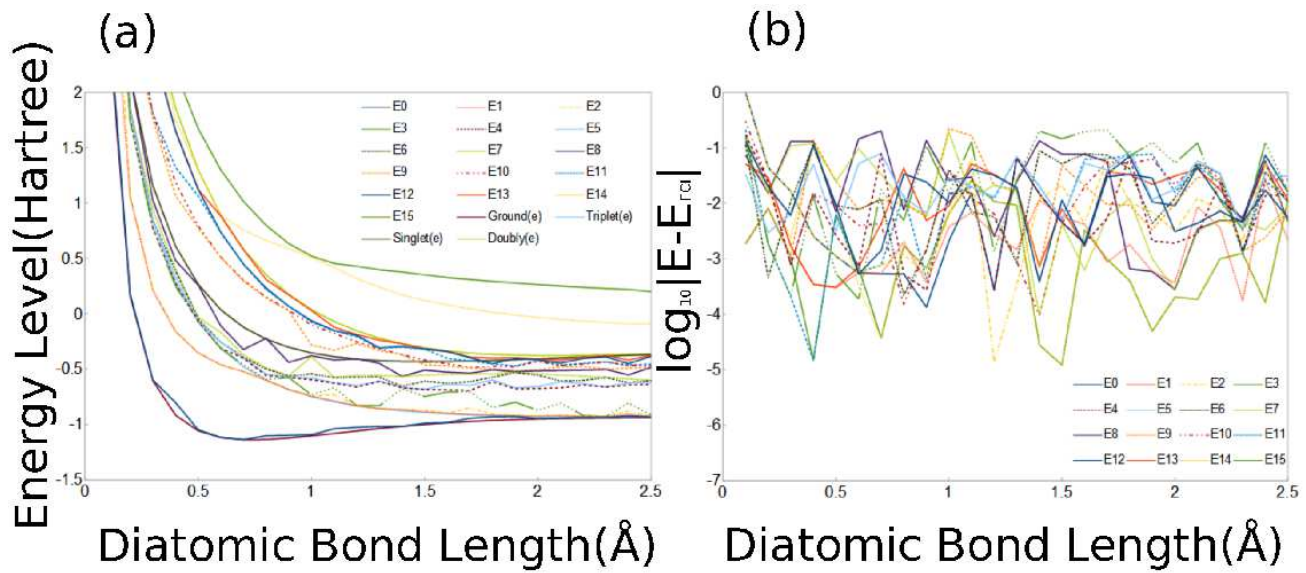}
\newline
\caption{Calculated energy levels(Hartree) of all 16 states on hydrogen molecule and their log error by VQSE with SSVQE-prescreening, respectively. (a) energy levels of each state for diatomic bond length r$(\AA)$. (b) log error($log_{10}\mid E - E_{FCI} \mid$) of each states. All data are sampled from 0.1 to 2.5 in 0.1 pitch.
}\label{VQSEps}
\end{figure*}

In this section, we show the result of VQSE with SSVQE pre-screening experiments. We set the number of iterations in SSVQE-pre-screening equal to 500. We display the calculated energy levels and the ordinary log of the difference between the calculated energies and exact value calculated by STO-3G classically (log error) for diatomic bond length from 0.1$(\AA)$ to 2.5$(\AA)$ in 0.1 pitch of ground, triplet, singlet and doubly excited states on hydrogen molecules by VQSE in Fig.\ref{VQSE}(A), (B). The more diatomic bond length increases, the less the accuracy of energy levels except those at the highest and second-highest levels. We show the calculated energy levels and the log errors for diatomic bond length from 0.1$(\AA)$ to 2.5$(\AA)$ in 0.1 pitch of ground, triplet, singlet and doubly excited states on hydrogen molecules by VQSE with SSVQE-prescreening in Fig. \ref{VQSEps}(a),(b). The accuracy of some states ares higher than that of VQSE without pre-screening in the range of $r \in [0.1,1.0]$. However, the accuracy of all energy levels is inferior to that of the VQSE with no pre-screening in case $r> 1.0$. The number of iterations of VQSE is not enough or SSVQE-pre-screening is not effective for the systems that all energy levels are near and highly degenerate.

\section{Comparison of number of iterations}\label{4}
\begin{figure*}
\includegraphics[scale=0.50]{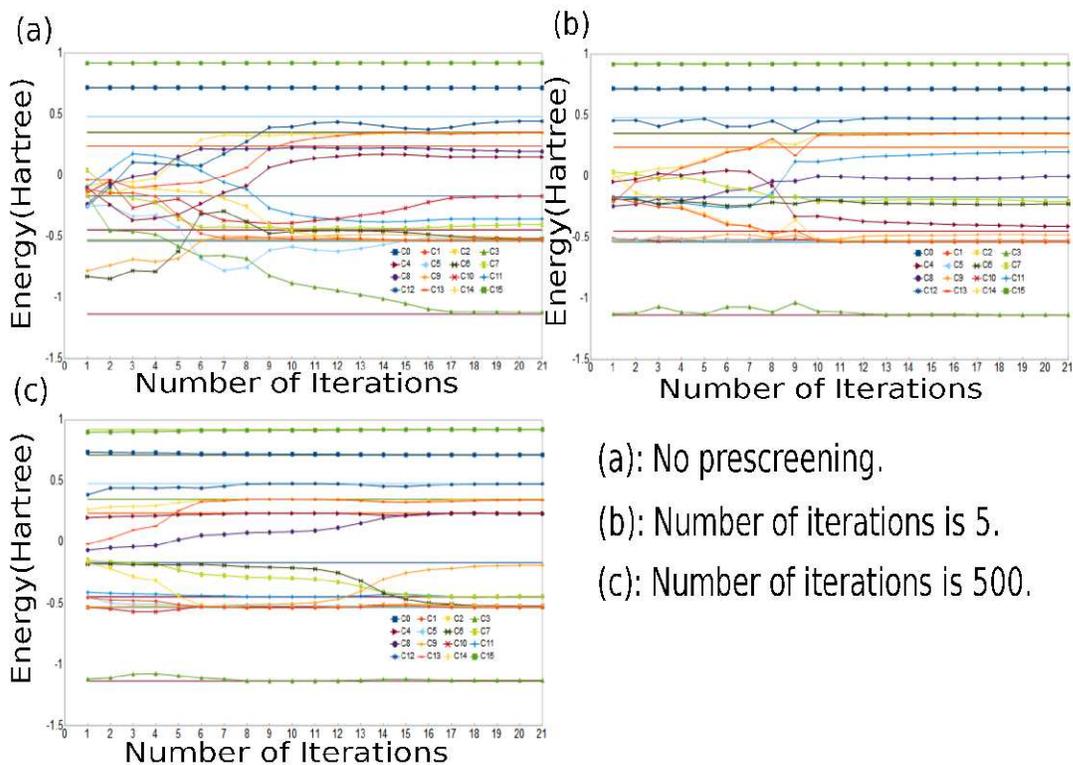}
\newline
\caption{The number of iterations v.s calculated energy levels(Hartree) of all 16 states on hydrogen molecule by VQSE with SSVQE-prescreening that the number of iterations is (a)0 (no prescreening), (b)5, and (c)500, respectively. 
}\label{Hconv}
\end{figure*}

The number of iteration of SSVQE-prescreening contributes to the accuracy of the VQSE method. In this section, we compare the convergence of energy levels of the number of iterations of SSVQE-pre-screening, which are 0, 5, and 500 on hydrogen and lithium hydride molecules. We show the result of convergence on hydrogen molecules in the case that the number of iterations are (a)0, (b)5, and (c)500 in Fig. \ref{Hconv}, respectively. We observe that the more the number of iterations are, the more energy levels converge close to the exact values. The difference between the calculated and exact values of 6 energy levels are less than 0.01 Hartree in case (b), and those between the calculated and exact values of 13 energy levels are less than 0.01 Hartree in case (c).

\begin{figure*}
\includegraphics[scale=0.50]{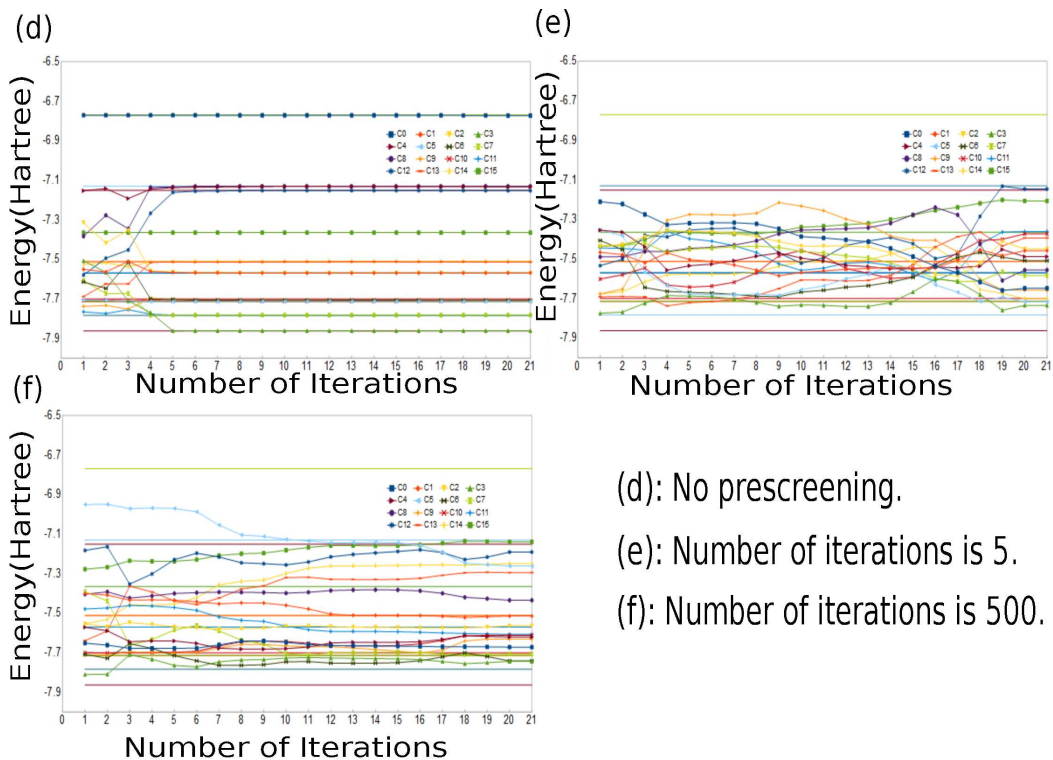}
\newline
\caption{The number of iterations v.s calculated energy levels(Hartree) of all 16 states on lithium hydride molecule by VQSE with SSVQE-prescreening that the number of iterations is (d)0 (no prescreening), (e)5, and (f)500, respectively. 
}\label{LiHconv}
\end{figure*}

In contrast, SSVQE-prescreening does harm to VQSE on lithium hydride molecules. We show the result of convergence on lithium hydride molecules in the case where the number of iterations is (d)0, (e)5, and (f)500 in Fig. \ref{LiHconv}, respectively. Almost all energy levels converged to exact energy levels in case (d) the number of iterations is 0(no pre-screening). However, only 6 and 4 energy levels converged to exact values in case (e) the number of iterations is 5 and (f)500, respectively. The differences between the calculated and exact values of 6 energy levels are less than 0.01 Hartree in both cases (e) and (f). The differences between the calculated and exact values of 8 energy levels are less than 0.01 Hartree in case (d).

According to the above data, in case all states converge to their exact values, more than two non degenerate energy levels almost matched with their exact values before the iteration on VQSE. Both H$_2$ and LiH have 11 degenerated energy levels in 5 states. Hence, close energy levels to converged non degenerate energy levels to exact them accelerated left energy levels converging to left degenerated energy levels.

\section{Concluding remarks}\label{5}
In this paper, it is confirmed that SSVQE-prescreening is not effective for the highly degenerated systems, and converging of non degenerate energy levels leads to convergence of left energy levels. VQSE is weak for degenerate states, thus, pre-screening must prepare the parameter sets that are close to the exact value of non degenerate states. This is the next problem to address.

However, only for the systems that have distant energy levels and non degenerate energy levels, it is confirmed that VQSE with SSVQE-prescreening contributes. VQSE with no prescreening has a low accuracy for some energy levels in such cases. High accuracy is necessary for the initial states of the Quantum Phase Estimation Algorithm. Hence, this algorithm can contribute to the era of coming near-term mid-scale quantum computers.

\newpage
\bibliographystyle{apsrev4-2}
\bibliography{main}

\end{document}